\documentstyle[prd,aps]{revtex}
\begin{document} 
\title{Pionic coupling constants of heavy mesons in the quark model} 
\author{Dmitri Melikhov$^{a,}$\footnote{On leave of absense from Nuclear
Physics Institute, Moscow State University, Moscow, 119899, Russia} and
Michael Beyer$^{b}$}
\address{$^a$LPTHE,
Universit\'e de Paris XI, B\^atiment 211, 91405 Orsay Cedex, France\\
$^b$Physics Department, Rostock University, D-18051 Rostock, Germany}
\maketitle
\begin{abstract} 
We analyse pionic couplings of heavy mesons combining PCAC with the
dispersion quark model to calculate the relevant transition form
factors. Ground states and radial excitations are considered. For the
ground state coupling constants the values $\hat g=0.5\pm0.02$ 
in the heavy quark limit, and $g_{B^*B\pi}=40\pm 3$, 
$g_{D^*D\pi}=16\pm 2$ are obtained. A sizeable suppression 
of the coupling constants describing the pionic decays of the radial 
excitations is observed. 
\end{abstract}
\vspace{0.5cm}
Pionic coupling constants of heavy mesons are basic quantities to 
understand the heavy meson lifetimes: they govern strong decays 
of heavy mesons by the emission of a pion which in practice turn out to 
be most important strong decay modes. 
The knowledge of the pionic coupling constants of the ground-state as well as 
the excited 
heavy mesons is timely in particular in view of the new experimental 
data on the excited heavy mesons \cite{delphi}. 

The heavy quark (HQ) symmetry provides important relations between 
the coupling constants of pionic transitions of various mesons 
containing a heavy quark. In the leading $1/m_Q$ order (LO) the heavy
quark spin decouples from other degrees of freedom \cite{hqs}. Thus  
in the limit of an infinitely heavy spectator quark, the respective 
pionic decay rates for mesons with different spins but 
same quantum numbers of the light degrees of freedom 
(total momentum of the light degrees of 
freedom $j$, angular momentum $L$, etc) \cite{iw2} are equal.   

Note, however that the heavy quark expansion which provides many 
rigorous results for {\it semileptonic} transitions between heavy 
mesons, turns out to be much less efficient in the case of the {\it strong}   
transitions. 
In fact, these two types of processes have quite 
different dynamics. Semileptonic transitions of heavy mesons are induced by 
the heavy quark weak transition and in this case HQET predicts not only the 
structure of the $1/m_Q$ expansion of the long-distance (LD) contributions to the 
form factors, but also provides important absolute 
normalization of the LO form factors. 
On the other hand, in the case of strong pionic decay the heavy quark
remains spectator and the dynamics of the process is determined by the 
light degrees of freedom. As a consequence HQ symmetry does not provide 
any absolute normalization, thus making the pionic
decays of heavy mesons a more complicated problem to analyse   
compared with semileptonic decays. 

The amplitudes of the strong pionic  
vector-to-vector and pseudoscalar-to-vector transitions have the structure  
\begin{eqnarray}
\langle V(p_2)\pi(q)|P(p_1)\rangle &=&q_\nu 
\varepsilon_\nu(p_2) g_{VP\pi}\nonumber\\
\label{v2vc}
\langle V(p_2)\pi(q)|V(p_1)\rangle &=&\varepsilon_\nu(p_1)\varepsilon_\mu(p_2)
p_{1\alpha}p_{2\beta}\epsilon^{\nu\mu\alpha\beta} g_{VV\pi},   
\end{eqnarray}
where $\varepsilon$ denotes the polarization vector of the vector meson. 
The coupling constants can be expanded in a $1/m_Q$ series as follows 
\begin{eqnarray}
\frac{f_\pi}{2\sqrt{M_PM_V}}g_{VP\pi}&\equiv&\hat g_{PV\pi}=
\hat g+\frac{\hat g^{(1)}_{P}}{m_Q}+\dots, 
\nonumber\\
\frac{f_\pi}{2}g_{VV\pi}&\equiv&\hat g_{VV\pi}=
\hat g+\frac{\hat g^{(1)}_{V}}{m_Q}+\dots, 
\end{eqnarray}
and due to the HQ symmetry in the leading $1/m_Q$ order both constants are 
governed by the same quantity $\hat g$, 
whereas the higher order terms are different. 

The pionic coupling constants for heavy mesons $g_{DD^*\pi}$ and $g_{BB^*\pi}$
have been analysed within various versions of the constituent quark model 
\cite{physrep}, sum rules \cite{sr} and using the lattice simulations \cite{lat}.  

The results of the quark model strongly 
depend on the particular version of the QM used: the values of $\hat g$ 
range from 1 (nonrelativistic quark model) to 1/3 
(the Salpeter equation with massless quarks) \cite{physrep}. 
LCSR \cite{sr} obtained a small value $0.32\pm0.02$. Recent lattice 
simulations reported 0.42 with however large statistical errors 
$\pm 0.04\pm 0.08$. Thus at the moment no reliable predictions are available 
even for the coupling constants of the ground state mesons. An analysis of the pionic 
decays of radially excited states is almost lacking. 
This calls for further investigation and a better understanding 
of the problem. 

Although the present QM results are strongly scattered, one feature of this
approach still seems to be very attractive: namely, various processes can be 
connected to each other via the wave functions of the mesons involved, 
once a proper formulation of the QM is used. We believe that 
such a proper formulation should (i) be based on a relativistic consideration 
and (ii) reproduce rigorous QCD results in the known limits, e.g. in the limit 
when the meson decay is induced by the HQ transition. 
The dispersion formulation of ref \cite{m1} was shown to satisfy these 
requirements. Recently, we have applied this dispersion 
approach to analyse the $B\to\pi$ transition form factors and 
determined the relevant quark model parameters and the wave function of the 
$B$ meson \cite{bm}. Once the whole framework and the numerical parameters 
are fixed we expect to perform a reliable analysis of the pionic coupling 
constants of the heavy mesons. An additional attractive feature of this 
approach is that it is straight forward to incorporate the excited states, 
in particular radial excitations. 

The dispersion approach is based on 
the spectral representations of the form factors through the wave functions of the
initial and final mesons. The double spectral densities are obtained from the 
relevant Feynman graphs, whereas the subtraction terms remain apriori ambiguous and 
should be determined from some other arguments. 
In ref \cite{m1} the subtraction terms have been determined such that the 
correct structure of the $1/m_Q$ expansion in the leading and subleading orders 
is reproduced. Such a procedure of determining the 
subtraction terms provides a reliable description of the form factors 
if the decaying quark is heavy for both heavy and light quarks produced in the 
weak decay. However such a strategy might be not efficient for 
the case when the interacting quark is light but the spectator is heavy, 
that is exactly the case of the pionic decay of a heavy meson, 
and requires proper modification. 

In this letter we analyse the pionic transitions of the ground state and radially 
excited heavy mesons, studying in parallel the $V\to V\pi$ and $V\to P\pi$ cases.
We make use the PCAC which allows one to reduce the 
calculation of the complicated amplitude involving three hadrons to a 
considerably more simple amplitudes of the $P\to V$ and $V\to V$ transitions induced by the 
light-quark axial-vector current. The corresponding amplitudes have the following structure 
\begin{eqnarray}
%\label{v2v}
\langle V(p_2)|A_\mu|V(p_1) \rangle &=&
i\varepsilon_\alpha(p_1)\varepsilon_\beta(p_2)
\epsilon_{\mu\nu\alpha\beta}P_\nu h(q^2)/2+\dots, \nonumber\\
\label{v2p}
\langle V(p_2)|A_\mu|P(p_1)\rangle&=&i\varepsilon_\nu(p_2)[g_{\mu\nu}f(q^2)+
a_+(q^2)P_\mu q_\nu+a_-(q^2)q_\mu q_\nu],
\end{eqnarray}
where $q=p_1-p_2$, $P=p_1+p_2$, and the dots denote terms transverse with respect 
to $q_\mu$ the particular form of which is not important for us. 
The form factors $a_-$ and $h$ contain a pole at $q^2=m_\pi^2$ due to the contribution of the 
intermediate pion state in the $q^2$-channel, and the residues in these poles are expressed through the pionic 
coupling constants $g_{VP\pi}$ and $g_{VV\pi}$ as follows
\begin{eqnarray}
h(q^2)&=&\frac{f_\pi q^2 g_{VV\pi}}{m_\pi^2-q^2}+\bar h(q^2) \nonumber\\
a_-(q^2)&=&\frac{f_\pi g_{VV\pi}}{m_\pi^2-q^2}+\bar a_-(q^2),   
\end{eqnarray}
where $\bar h$ and $\bar a_-$ are regular at $q^2=m_\pi^2$. 

In the $V\to V$ case, we find that the function $\Phi_V(q^2)\equiv h(q^2)(m_\pi^2-q^2)$ 
is smooth between the point $q^2=0$ and $q^2=m_\pi^2$ and due to the smallness 
of the pion mass we assume that $\Phi_V(0)\simeq \Phi_V(m_\pi^2)$. 
This yields the relation 
\begin{equation}
\label{gvvpi}
g_{VV\pi}=h(0)/f_\pi. 
\end{equation}
For the $P\to V$ case an additional step is necessary: taking first the 
divergence of the axial-vector current we find   
\begin{eqnarray}
\nonumber  
<V(p_2)|q_\mu A_\mu|P(p_1)>=q_\nu\epsilon_\nu(p_2)\frac{\Phi_P(q^2)}{m_\pi^2-q^2} 
\end{eqnarray}
with  
$\Phi_P(q^2)=\left[f(q^2)+(p_1^2-p_2^2)a_+(q^2)+q^2 \bar a_-(q^2)\right](m_\pi^2-q^2)
+q^2 f_\pi g_{VP\pi}$. 
A standard PCAC assumption $\Phi_P(m_\pi^2)\simeq \Phi_P(0)$ yields 
\begin{equation}
\label{g2}
g_{PV\pi}=\frac{1}{f_\pi}[f(0)+(p_1^2-p_2^2)a_+(0)]. 
\end{equation}
Notice that the pion should not be necessarily soft and the relations (\ref{gvvpi}) and 
(\ref{g2}) can be readily applied to the case when the masses $M_1$ and $M_2$ 
are substantially different as e.g. for the transition between the ground-state 
and radially-excited heavy mesons. 

The relations (\ref{gvvpi}) and (\ref{g2}) represent the coupling constants of interest 
through the transition form factors, for which we apply the dispersion approach. 

\section{Spectral representations of the coupling constants} 
We start with considering the spectral representation of the form factors 
along the lines of ref \cite{m1}. We apply this method to the case of the heavy 
spectator quark and propose proper modifications to calculate the pionic coupling constants. 

The dispersion approach gives the transition form
factors of the meson $M_1$ to the meson $M_2$ 
as double relativistic spectral representations in terms of 
the soft wave functions 
$\psi_1(s_1)=G_1(s_1)/(s_1-M_1^2)$ and $\psi_2(s_2)=G_2(s_2)/(s_2-M_2^2)$, 
of the initial and final mesons, respectively.  
To be specific, in the case of the weak decay $P_Q\to V_{q'}$ induced by the 
weak quark transition $Q\to q'$ where $q'$ might be both heavy and light 
the form factor can be represented as follows \cite{m1} 
\begin{eqnarray}
\label{dr}
\nonumber
f(q^2)=\int \frac{ds_1\;G_1(s_1)}{s_1-M_1^2}\;\frac{ds_2\;G_2(s_2)}{s_2-M_2^2} 
&&\left[\tilde f_D(s_1,s_2,q^2)+
\left((s_1-M_1^2)+(s_2-M_2^2)\right)
\tilde g_D(s_1,s_2,q^2)\right.\\
&&\left.+(s_1-M_1^2)\xi_1(s_1,s_2)+(s_2-M_2^2)\xi_2(s_1,s_2)\right], 
\end{eqnarray}
where $\tilde f_D={\rm disc}_{s_1}{\rm disc}_{s_2} f_D(s_1,s_2,q^2)$ 
and $\tilde g_D={\rm disc}_{s_1}{\rm disc}_{s_2} g_D(s_1,s_2,q^2)$ 
are directly calculated from the triangle Feynman graph with pointlike vertices as 
double spectral densities of the relevant form factors.  
The functions $\xi_1$ and $\xi_2$ are not known precisely but are 
known to behave as $1/m_Q$ in the limit $m_Q\to\infty$, where $m_Q$ is the 
mass of the initial heavy quark. It is important to point out at
the hierarchy of different terms in eq (\ref{dr}) in case the 
initial active quark is heavy: namely, the term proportional to 
${\rm disc}_{s_1}{\rm disc}_{s_2} f_D$ contributes in the LO 
(and all other orders as well), the subtraction term proportional to 
${\rm disc}_{s_1}{\rm disc}_{s_2} g_D$ contributes starting from the
subleading order, and the subtraction terms proportional to $\xi$ contribute only in the higher 
orders. So this expansion works well if the initial active quark is heavy but the 
spectator quark is light, both for the cases of the light and heavy active quark in the final
state. 

However in case the spectator is heavy and the active quark 
is light, the subtraction term proportional to $\tilde g_D$ as well as the 
functions $\xi$ contain powers of the spectator mass, and all terms have the same 
order of magnitude and thus should be considered on an equal footing. Hence, the
expansion (\ref{dr}) becomes ineffective in practice and 
the procedure requires a modification. 

To find such proper modification it is convenient to start with considering 
the $V_1\to V_2$ transition induced by the axial-vector current. 
In the quark model the process is represented by the triangle graph
with relevant vertices describing the quark structure of the vector 
meson states. The procedure described in detail in  
\cite{m1} yields for the $g_{V_1V_2\pi}=h(q^2=0)/f_\pi$ the following 
spectral representation  
\begin{eqnarray}
\label{gv2}
g_{V_1V_2\pi}&=&\frac{m_1}{4\pi^2 f_\pi}
\int\limits_{(m_1+m_3)^2}^{\infty} ds \Psi_{V_1}(s)\Psi_{V_2}(s)\nonumber\\
&&\times\left[
m_1\log(r)+(m_3-m_1)\frac{\lambda^{1/2}}{s} +\frac{1}{\sqrt{s}+m_1+m_3}
\left(\frac{s+m_1^2-m_3^2}{s}\lambda^{1/2}-\frac{m_1^2}{2}\log(r)\right)\right], 
\end{eqnarray}
with $r=(s+m_1^2-m_3^2+\lambda^{1/2})/(s+m_1^2-m_3^2-\lambda^{1/2})$,  
$\lambda=(s-m_1^2-m_3^2)^2-4m^2_1m^2_3$, and $m_1$ and $m_3$ the masses 
of the active and the spectator constituent quarks, respectively 
(we follow the notations of \cite{m1}). 

This spectral representation can be rewritten in a more conventional 
form as an integral over the light-cone variables as follows 
\begin{eqnarray}
\label{gv1}
g_{V_1V_2\pi}=\frac{m_1}{2\pi^2 f_\pi}\int \frac{dxdk^2_\perp}{x(1-x)^2}
\Psi_{V_1}(s)\Psi_{V_2}(s)\left[m_1 x+m_3(1-x)
+\frac{2k^2_\perp}{\sqrt{s}+m_1+m_3}\right].  
\end{eqnarray} 
Here $s=\frac{m_1^2}{1-x}+\frac{m_3^2}{x}+\frac{k^2_\perp}{x(1-x)}$, and 
$x$ and $1-x$ are the fraction of the light-cone momentum carried by the spectator 
and the active quark, respectively. 

The expression (\ref{gv2}) is obtained as a double dispersion 
representation in the invariant masses of the initial and final $q\bar q$ 
pairs, the spectral density of which is calculated from the
triangle Feynman graph. At $q^2=0$ it is reduced to a simple form (\ref{gv2}). 
In principle subtraction terms can be added to this double spectral
representation. However, in the case of the $V\to V$ transitions 
there are no reasons dictating a necessity of subtractions, and we assume that 
the subtraction terms are absent. 

The representation (\ref{gv2}) describes the $\pi_0\to 2\gamma$ 
decay if we set $m_1=m_3$ and take the quark-photon vertex in the form 
$\bar q\gamma_\mu q$. In this case only the first term in
(\ref{gv2}) is present, and the photon wave function corresponding to
the pointlike interaction reads $\Psi_\gamma(s)=1/s$. 
So, $g_{\gamma\gamma\pi}$ becomes independent of the quark mass and we 
simply reproduce the value of the axial anomaly \cite{abj} 
from the imaginary part of the triangle graph (cf \cite{zakharov} and refs 
therein).  

The axial-vector current satisfies the equation of motion 
$\partial_\mu A_\mu=2m j_5$, so the coupling constant $g_{V_1V_2\pi}$ 
can be equivalently calculated directly from the amplitude
$\langle V_2(p_2)|2mj_5(0)|V_1(p_1) \rangle=
\varepsilon_\alpha(p_1)\varepsilon_\beta(p_2)
\epsilon_{\mu\nu\alpha\beta}q_\nu P_\nu h(q^2)/2$. 

We proceed the same way to analyse the $P\to V$ transition: 
instead of calculating the full representations for $f$ and $a_+$ and then 
taking their linear combination (\ref{g2}), we focus on the amplitude 
$\langle V(p_2)| 2mj_5| P(p_1)\rangle$. Similar to the $V\to V$ case, 
this amplitude provides the correct double spectral density of the spectral representation 
for $g_{VP\pi}$ which takes the form  
\begin{eqnarray}
\label{gp1}
g_{VP\pi}&=&\frac{m_1}{4\pi^2 f_\pi}\int\limits_{(m_1+m_3)^2}^{\infty}
ds\;\Psi_P(s)\Psi_V(s)         \nonumber \\
&&\times\left[\left(s-(m_1-m_3)^2\right)\log(r)
-\left(1+\frac{2m_1}{\sqrt{s}+m_1+m_3}\right)
\left((s+m_1^2-m_3^2)\log(r)-2\lambda^{1/2}\right)\right] \nonumber \\ 
&=&\frac{m_1}{4\pi^2 f_\pi}\int\frac{dxdk^2_\perp}{x(1-x)^2}
\Psi_P(s)\Psi_V(s)\left[
s-(m_1-m_3)^2 -\frac{k^2_\perp}{1-x}\left(1+\frac{2m_1}{\sqrt{s}+m_1+m_3}\right)
\right]. 
\end{eqnarray}
The form of the relations (\ref{g2}) and (\ref{dr}) however prompts that 
in distinction to the $V\to V$ transition, in the 
$P\to V$ transition the subtraction term is nonzero. The latter cannot be 
determined uniquely within the dispersion approach. 
It is very important however that the HQ symmetry ensures the 
subtraction term to contribute only in the subleading $1/m_Q$ order. 
Namely, the HQ symmetry predicts the LO relation between the coupling constants 
$g_{VP\pi}=m_Q g_{VV\pi}$. The double spectral densities of the 
representations for $g_{VV\pi}$ (\ref{gv1}) and $g_{VP\pi}$ (\ref{gp1}) satisfy 
this relation. Hence the subtraction terms in $g_{VV\pi}$ and $g_{VP\pi}$
should also have the same LO $1/m_Q$ behavior. Since the subtraction 
term in $g_{VV\pi}$ is absent, the subtraction term in $g_{VP\pi}$ does not  
contribute in the LO. Although we cannot determine the subtraction term 
uniquely, several reasonable ways of fixing this subtraction term yield a 
numerical uncertainty in $g_{PV\pi}$ to be not more than 10\%. 
 
The normalization condition for the radial wave functions $\Psi$ of the ground state 
and the radial excitation of the vector and the pseudoscalar mesons has the form 
\begin{eqnarray}
\label{norm1}
\frac{1}{8\pi^2}\int ds\;\Psi_i(s)\;\Psi_j(s)\;\frac{\lambda^{1/2}}{s}[s-(m_1-m_3)^2]
=\frac{1}{8\pi^2}\int \frac{dxdk^2_\perp}{x(1-x)}\Psi_i(s) \Psi_j(s)[s-(m_1-m_3)^2]
=\delta_{ij}. 
\end{eqnarray}

It is easy to see that in the nonrelativistic (NR) limit 
$|\vec k|=\lambda^{1/2}/2\sqrt{s}\ll m_1,m_3$ the coupling constants take the values 
$g^{NR}_{PV\pi}=2M/f_\pi$ and $g^{NR}_{VV\pi}=2/f_\pi$. Moreover, in the NR limit 
the coupling constants of the transition between the ground state and the radial 
excitation vanish, i.e. $g^{NR}_{PV'\pi}=g^{NR}_{VV'\pi}=0$. 
 
For the analysis of the HQ expansion it is convenient to introduce a 
new variable $z$ such that $s=(z+m_1+m_3)^2$ and use the fact that the soft wave
functions $\phi_0$ are localized in the region $z\simeq \Lambda_{QCD}$. 
Performing the HQ expansion of all quantities including $\Psi$ 
in the inverse powers of $m_3=m_Q\to \infty$ and keeping $m_1=O(\Lambda_{QCD})$ 
(see \cite{m1}) we find 
\begin{eqnarray}
\label{hatg}
\hat g_{ij}=\frac{1}{2}\int dz \phi^{i}_0(z)\phi^{j}_0(z)
m_1\left[
m_1\log\left(\frac{z+m_1+\sqrt{z(z+2m_1)}}{z+m_1-\sqrt{z(z+2m_1)}}\right)
+2\sqrt{z(z+2m_1)}\right].
\end{eqnarray}
In this expression $\phi^{i,j}_0$ are the LO soft wave function of the 
ground state and the radially excited mesons which satisfy the 
normalization condition 
\begin{eqnarray}
\int dz \phi^{i}_0(z)\phi^{j}_0(z)\sqrt{z}(z+2m_1)^{3/2}=\delta_{ij}
\end{eqnarray}
Notice that the Isgur-Wise function is directly expressed through 
$\phi_0$ (see \cite{m1}).  
These formulas provide a possibility to evaluate the coupling constants of interest 
if the numerical parameters of the model are known.

\section{Numerical analysis} 
We now provide the numerical estimates of the LO quantity $\hat g$ for the ground-state
and radially excited mesons. The spectral representation for this
quantity is completely fixed and we need to choose the proper numerical parameters
of the model. As found in many applications of the dispersion approach 
(see \cite{m1,bm} and refs therein), 
an approximation of the soft wave function by an exponent provides reasonable estimates. 
Assuming the wave function to have the form 
$\Psi(s)\simeq \exp(-\vec k^2/2\beta^2)$, the LO radial 
wave function of the ground state reads \cite{m1} 
\begin{eqnarray}
\phi(z)\simeq\sqrt{\frac{z+m_1}{z+2m_1}}\exp\left(-\frac{z(z+2m_1)}{2\beta^2_\infty}\right).
\end{eqnarray}
Similarly, we assume for the wave function of the radial excitation the form 
\begin{eqnarray}
\phi_r(z)\simeq\left({1-C_r\frac{z(z+2m_1)}{2\beta^2_\infty}}\right)
\exp\left(-\frac{z(z+2m_1)}{2\beta^2_\infty}\right), 
\end{eqnarray}
where the normalization factor and the coefficient $C_r$ are fixed by the normalization
conditions and the orthogonality between the wave functions of the ground state and the
radial excitation. 

The light-quark mass was determined from the numerical matching of the transition 
form factors in the $B\to\pi,\rho$ decay to the available lattice data
\cite{latff} 
and was found to
be tightly restricted to the range $m_1=0.23\pm0.01\; GeV$. Therefore the only unknown 
parameter is $\beta_\infty$. The analysis of the leptonic decay constants of 
heavy mesons shows that simple approximate relations  
\begin{eqnarray}
\beta_P(m_Q)=\beta_\infty(1-C_P/m_Q), \qquad 
\beta_V(m_Q)=\beta_\infty(1-C_V/m_Q) 
\end{eqnarray}
with $C_P\simeq 0.1$, $C_V\simeq 0.2$ and $\beta_\infty=0.5$ yield the values of $\beta$ 
which provide reasonable leptonic constants of the charm and beauty mesons 
(see Table 1) calculated within the quark model through formulas given in \cite{m1,jaus}.  
The slope parameter of the IW function for such $\beta_\infty$ is 
$\rho^2=1.2\pm0.03$ \cite{m1}. 
The eq (\ref{hatg}) then yields 
\begin{equation}
\label{reshatg}
\hat g=0.5, \qquad \hat g_r=0.11.
\end{equation}
Thus for the LO quantity the orthogonality of the radial wave functions provides 
a strong suppression of $\hat g_r$\footnote{Notice that in contrast to the PCAC based 
approaches, the NR quark-model calculations based on taking into account 
the quark structure of the emitted pion through the pion wave function yield bigger 
values of $\hat g$ and do not yield a suppression of $\hat g_r$ since the product 
of the orthogonal wave function is smeared by the integration 
with the pion wave function and as the result the orthogonality 
is not efficient \cite{pene}.}. 

The results of calculating the pionic coupling constants of the $B$ and $D$ 
mesons are given in Table 2. Whereas the $g_{VV\pi}$ is determined quite reliably 
within the dispersion apprach, 
the $g_{PV\pi}$ suffers from the intrinsic 
uncertainty of the approach based on the impossibility to fix the subtraction terms. 
Assuming several reasonable choices of  
subtraction terms in the form factors $f$, $a_+$ and $g$ which provide the correct
behaviour in the limit of the heavy active quark, but differ for the case of the 
heavy spectator we have found that in all cases, the difference in the coupling 
constants of the ground state is not more than $10\%$. 
This minor difference is easily understood taking into account that the subtraction 
terms contribute only in the subleading order. 

Notice that there is a possibility of an alternative determination of the $g_{B^*B\pi}$ 
from the form factors of the semileptonic $B\to\pi$ transition near zero recoil. Namely, 
\begin{eqnarray}
f_0(M_B^2)\simeq f_0(M_B^{*2})=\frac{g_{B^*B\pi}f_{B^*}}{2M_{B^*}}(1+r'(1)), 
\end{eqnarray}
where $r(\hat q^2=q^2/M^2_{B^*})$ is defined as follows 
\begin{eqnarray}
r(\hat q^2)=-\frac{f_-(q^2)}{f_+(q^2)}\frac{M^2_{B^*}}{M_B^2-M_\pi^2}, 
\end{eqnarray}
The function $r(\hat q^2)$ satisfies the condition $r(1)=1$ and 
is known numerically from the results of the lattice simulations in the 
region $\hat q^2\le 0.7$
\cite{latff}, in particular $r(0.7)\simeq 0.9$. 
The function $r(\hat q^2)$ is regular near $\hat q^2=1$ and thus there 
are no reasons for fast variations of this function near $\hat q^2=1$. 
Then, taking into account the lattice data we can estimate $r'(1)\le 0.8$. 
Using the current algebra relation 
$f_0(M_B^2)=f_B/f_\pi$, neglecting the $1/m_b$ corrections which are 
small, and taking the ratio of the leptonic constants 
$f_{B^*}/f_B\simeq 1.2$ we find  
\begin{eqnarray}
\hat g=\frac{1}{f_{B^*}/f_B(1+r'(1))}\ge 0.45  
\end{eqnarray} 
in agreement with (\ref{reshatg}) and with the recent result of the lattice
simulation \cite{lat}.  
We would like to point out 
that the value $\hat g\simeq 0.27$ proposed in \cite{stewart} as the
preferrable solution found from the description of the $D^*$ decays falls 
far out of this estimate. 

The calculation of the coupling constants of the pionic
transition between the ground state and the radial excitation is more involved. 
In this case the LO term is itself strongly suppressed because of the orthogonality 
of the wave functions. So the dependence on the parameters of the wave function of 
the radial excitation in the case of the $g_{VV'\pi}$ and, in addition to this, also 
the dependence on the particular choice of the subtraction procedure in $g_{PV'\pi}$ 
is more sizeable. Table 2 presents numerical results. 
The slope parameter of the Gaussian wave function of the radial excitation 
at finite masses is not known, so for estimating $g_{VV'\pi}$ 
we assume that this parameter lies within a 10\% interval around the corresponding 
ground-state slope. 
In the case of $g_{PV'\pi}$ the relation (\ref{g2}) prompts that the $1/m_Q$ corrections 
numerically might be sizeable if the $M_{V'}^2-M_P^2$ is big as it is for the radial 
excitations of heavy mesons. In this case we cannot obtain a reliable estimate
and Table 2 provides the lower bounds found from the spectral representations without 
subtractions. 
\section{Conclusion}
We have analysed the pionic coupling constants of heavy mesons within the 
framework based on the combination of PCAC with the 
dispersion approach and obtained the following results:

1. Spectral representations of the coupling constants in terms of the  
wave functions of the initial and final heavy mesons have been obtained.  
The double spectral densities of these spectral representations for  
$g_{PV\pi}$ and $g_{VV\pi}$ are equal in the HQ limit in agreement 
with the HQ symmetry. 

In the case of the $V\to V$ transition the calculation of the $g_{VV\pi}$ is
equivalent to the calculation of the amplitudes $\langle V|2mj_5|V\rangle$. 
There is no reason dictating a necessity of any subtraction term in the 
spectral representation for $g_{VV\pi}$ and thus the latter is determined 
unambiguously within our approach. 

In the $P\to V$ case the double spectral density of $g_{VP\pi}$ can be also 
found from the amplitude $\langle V|2mj_5|P\rangle$. 
The spectral representation for $g_{VP\pi}$, however, contains a subtraction term 
which cannot be fixed within our approach. 
Important is that due to the HQ symmetry this subtraction term does not 
contribute in the leading $1/m_Q$ order\footnote{Strictly speaking, 
we cannot completely exclude a possibility of 
subtraction terms both in $\hat g_{VV\pi}$ and $\hat g_{PV\pi}$ which give the same 
contributions to both quantities in the LO. However, unless no reasons for 
such subtraction terms are found we do not include them into consideration. 
In a recent analysis of the $B\to \pi$ form factors \cite{bm} we have found 
that the behavior of $f_0$ at zero recoil is compatible with 
$\hat g_{B^*B\pi}\simeq 0.5-0.7$. The values of $\hat g_{B^*B\pi}$ as big as 
$0.6-0.7$ can be obtained only by assuming the subtraction terms both in 
$g_{B^*B\pi}$ and $g_{B^*B^*\pi}$ which give a nonvanishing contribution 
already in the leading $1/m_Q$ order.} and thus numerically is not essential 
at least for the $g_{B^*B\pi}$. 

2. A spectral representation for the LO $1/m_Q$ coupling constant $\hat g$ which 
describes the pionic transition in the HQ limit has been obtained.  
The $\hat g$ is represented through the LO wave functions of the heavy mesons 
which determine also the Isgur-Wise function. The $\hat g$ 
is determined quite reliably within the dispersion approach since it is not 
affected by the uncertainties in the subtraction procedure. 

For the transition between the ground-state heavy mesons we have found  
$\hat g=0.5\pm 0.02$. We argue that the behavior of the $B\to\pi$ transition 
form factors near zero recoil provide independent arguments in favour of such 
considerably large value of $\hat g$. 

For the pionic transition between the ground state and the radial excitation 
in the HQ limit a strong suppression $\hat g_r=0.11\pm0.02$ is observed as a 
consequence of the orthogonality of the corresponding radial wave 
functions. 

3. Using the QM parameters obtained from the analysis of the $B\to\pi$ decay 
we have calculated the pionic coupling constants of the ground state charm and 
bottom mesons. Our final numerical estimates are listed in Table 2. 

For the $V\to V$ transition higher-order $1/m_Q$ effects in $g_{VV\pi}$ 
are found to be small and not to exceed 5-6\%. 

In the $P\to V$ case, the coupling constant $g_{PV\pi}$ depends on the choice of 
the subtraction procedure which cannot be fixed unambiguously. However, numerically 
the uncertainty estimated by using several reasonable subtraction 
procedures is not more than 10\%. 
 
4. We estimated the pionic couplings of the radially excited states 
$B^{*'}$ and $D^{*'}$. The errors turn out to be much bigger compared with the 
corresponding ground states: the LO contribution is strongly suppressed 
and thus the $1/m_Q$ effects are found to be numerically more important. 
As a result the coupling constants are sensitive to the specific form of the 
radial-excitation wave function and $g_{PV'\pi}$, in addition to this, 
strongly depends on the the details of the subtraction procedure. 
Therefore only lower bounds on $g_{PV'\pi}$ are given. 

\section{Acknowledgements} We are grateful to D. Becirevic, 
A. Le Yaouanc and O. P\'ene for stimulating discussions of this subject. 
The work was supported in part by DFG under grants 436 RUS 18/7/98 and 436 RUS 17/26/98, and 
by the NATO Research Fellowships Program. 
%\newpage

\begin{table}[1]
\caption{\label{table:parameters}
Quark masses and the slope parameters of the soft meson 
wave functions and the calculated leptonic decay constants (in GeV)} 
\centering
\begin{tabular}{|c|c|c||c|c|c|c|c||c|c|c|c|}
%\hline
$m_b$ & $m_c$ & $m_u$ & $\beta_\infty$ & $\beta_{B}$ & $\beta_{B^*}$ & $\beta_{D}$ & $\beta_{D^*}$  
& $f_{B}$ & $f_{B^*}$ & $f_{D}$ & $f_{D^*}$ \\ 
\hline
4.85 & 1.4 & 0.23 & 0.5 & 0.49 & 0.49 & 0.46 & 0.43  & 0.16 & 0.175 & 0.2 &0.22   
%\\
%\hline
\end{tabular}
\end{table}

\begin{table}[2]
\caption{\label{table:constants}
Pionic coupling constants of heavy mesons. The error bars correspond to the
variations of the slope parameters $\beta$ around  
the average values of Table 1 yielding a 10\% 
variations of the leptonic decay constants.} 
\centering
\begin{tabular}{|c|c|c|c|c|}
%\hline
       & $\hat g_{VV\pi}$ & $\hat g_{VP\pi}$ & $\hat g_{V'V\pi}$ & $\hat g_{V'P\pi}$ \\ 
\hline
$D,D^*$   &0.53$\pm$0.03 & 0.53$\pm$0.05 & 0.15$\pm$0.03  & $>0.14$\\  
\hline
$B,B^*$   & 0.5$\pm$0.02 & 0.5$\pm$0.04  & 0.12$\pm$0.03  & $>0.12$\\  
\hline
HQ-limit  & 0.5$\pm$0.02 &  $-$ &   0.11$\pm$0.02  & $-$  
%\\
%\hline
\end{tabular}
\end{table}
\end{document}